\documentclass[9pt,twocolumn,twoside]{my-pnas-arxiv}

\templatetype{pnasresearcharticle} 

\usepackage{graphicx}
\usepackage{subcaption}
\usepackage{braket}
\usepackage{xcolor}
\usepackage{bbold}
\usepackage{diagbox}

\def\psii{\psi_i}
\def\psif{\psi_f}
\def\id{\mathbb{1}}

\def\S#1{{\{#1\}_s}}
\def\S{\mathcal{S}}
\def\Ob{\mathcal{O}}
\def\ps{\mathrm{ps}}

\DeclareMathOperator{\Real}{Re}
\DeclareMathOperator{\Imag}{Im}
\DeclareMathOperator*{\E}{\mathbb{E}}
\DeclareMathOperator{\ad}{ad}
\newcommand*{\ketbra}[2]{\lvert #1 \rangle\!\langle #2 \rvert}

\def\L{\mathcal{C}}
\def\R{\mathcal{A}}
\def\PL{{\Pi_\L}}
\def\PLL#1{{\Pi^{#1}_{\L\L}}}

\def\PR{{\Pi_\R}}

\def\PRR#1{{\Pi^{#1}_{\R\R}}}
\def\PS#1{{\Pi^{#1}_{\mathcal{S}}}}

\begin{document}

\title{Variable-strength non-local measurements reveal quantum violations of classical counting principles}

\author[a,1]{Noah Lupu-Gladstein}
\author[a]{Ou Teen Arthur Pang}
\author[a]{Hugo Ferretti}
\author[a]{Weng-Kian Tham}
\author[a,b]{Aephraim M. Steinberg}
\author[a,c]{Kent Bonsma-Fisher}
\author[a]{Aharon Brodutch}

\affil[a]{Department of Physics and Center for Quantum Information and Quantum Control, University of Toronto, 60 St George St, Toronto, Ontario, M5S 1A7, Canada}
\affil[b]{Canadian Institute for Advanced Research, Toronto, Ontario, M5G 1M1, Canada}
\affil[c]{National Research Council of Canada, 100 Sussex Dr, Ottawa, Ontario, K1A 0R6, Canada}

\leadauthor{Lupu-Gladstein}

\significancestatement{Quantum theory has proven wildly successful in predicting properties of systems whose past or future are specified. Applying the theory to systems with a definite past \textit{and} future yields infamously counter-intuitive predictions e.g., three quantum pigeons can apparently occupy two pigeonholes without any pair occupying the same pigeonhole. Are such counter-intuitive predictions merely an artifact of measurement disturbance? We answer this question empirically by measuring photonic ``pigeons'' with a variety of different measurement disturbances by implementing a novel scheme that achieves variable-strength measurements of non-local observables. We discover that measurement disturbance can explain some, but not all, of the paradoxical phenomena of pre and postselected systems, thus revealing a truly strange quantum reality.}

\authorcontributions{N.L.G. built the experiment, collected and analyzed data, and wrote the paper. O.T.A.P., H.F., W.H.T., and K.B.F. assisted with data collection. H.F. and A.B. designed experiment. W.H.T. built photon source. A.B. helped write. A.M.S. supervised.}
\authordeclaration{The authors declare no competing interest.}
\correspondingauthor{\textsuperscript{2}To whom correspondence should be addressed. E-mail: nlupugla@physics.utoronto.ca}


\begin{abstract}
    We implement a variant of the quantum pigeonhole paradox thought experiment to study whether classical counting principles survive in the quantum domain. We observe strong measurements significantly violate the pigeonhole principle (that among three pigeons in two holes, at least one pair must be in the same hole) and the sum rule (that the number of pigeon pairs in the same hole is the sum of the number of pairs across each of the holes) in an ensemble that is pre and postselected into particular separable states. To investigate whether measurement disturbance is a viable explanation for these counter-intuitive phenomena, we employ the first ever variable-strength measurement of a non-local observable. As we decrease the measurement strength, we find the violation of the sum rule decreases, yet the pigeonhole principle remains violated. In the weak limit, the sum rule is restored due to the cancellation between two weak values with equal and opposite imaginary parts. We observe the same kind of cancellation at higher measurement strengths, thus raising the question: do strong measurements have imaginary parts?
\end{abstract}

\dates{This manuscript was compiled on \today}
\doi{\url{www.pnas.org/cgi/doi/10.1073/pnas.XXXXXXXXXX}}

\maketitle
\thispagestyle{firststyle}
\ifthenelse{\boolean{shortarticle}}{\ifthenelse{\boolean{singlecolumn}}{\abscontentformatted}{\abscontent}}{}

\firstpage[1]{4}

\dropcap{T}he correspondence between quantum measurements and physical reality  has been a source of intense debates since the early days of quantum theory \cite{EPR,Frauchiger2018,Pusey2012,Hardy2013,Mermin_2019}. One point of contention involves the connection between mathematically defined observables and our classical intuition about what these observables represent. The problem manifests starkly in \emph{pre and postselection}  (PPS) experiments where a system is first prepared in some initial state $\ket{\psii}$ (the preselection), then measured at some intermediate time, then projected onto some final state $\ket{\psif}$ (the postselection). Naively interpreting the result of the intermediate measurement as reflecting elements of physical reality can lead to paradoxes  \cite{Aharonov2005,Aharonov2002,Aharonov2013,Yokota2009,Hardy1992,BCsequential,Lundeen2009,Leifer2005,Pusey2015,Ravon2007,Resch2004,Aharonov2016,Aharonov2013b}. One resolution invokes measurement back-action, implying a disconnect between measurement results and the real state of affairs.

We present an experiment that probes the role of back-action in these paradoxes by varying the type and strength of measurement disturbance. The experiment is inspired by the so-called quantum pigeonhole paradox  \cite{Aharonov2016, reznik2020footprints}. The paradox arises from a counter-intuitive prediction that three pigeons placed among two holes can each occupy a different hole. This apparent logical contradiction has been observed indirectly at the weak limit in the correlations of a neutron interferometer \cite{Waegell2017} and directly at the the strong limit, using the Hong-Ou-Mandel effect to check if a pair of photons are in the same polarization state \cite{Chen2019}. Our experiment is the first to investigate the pigeonhole paradox across the full spectrum of measurement strengths, and indeed, the first to measure any non-local observable of a spatially distributed system over such a range of strengths.

\begin{figure}
    \centering
    \includegraphics[width=\columnwidth]{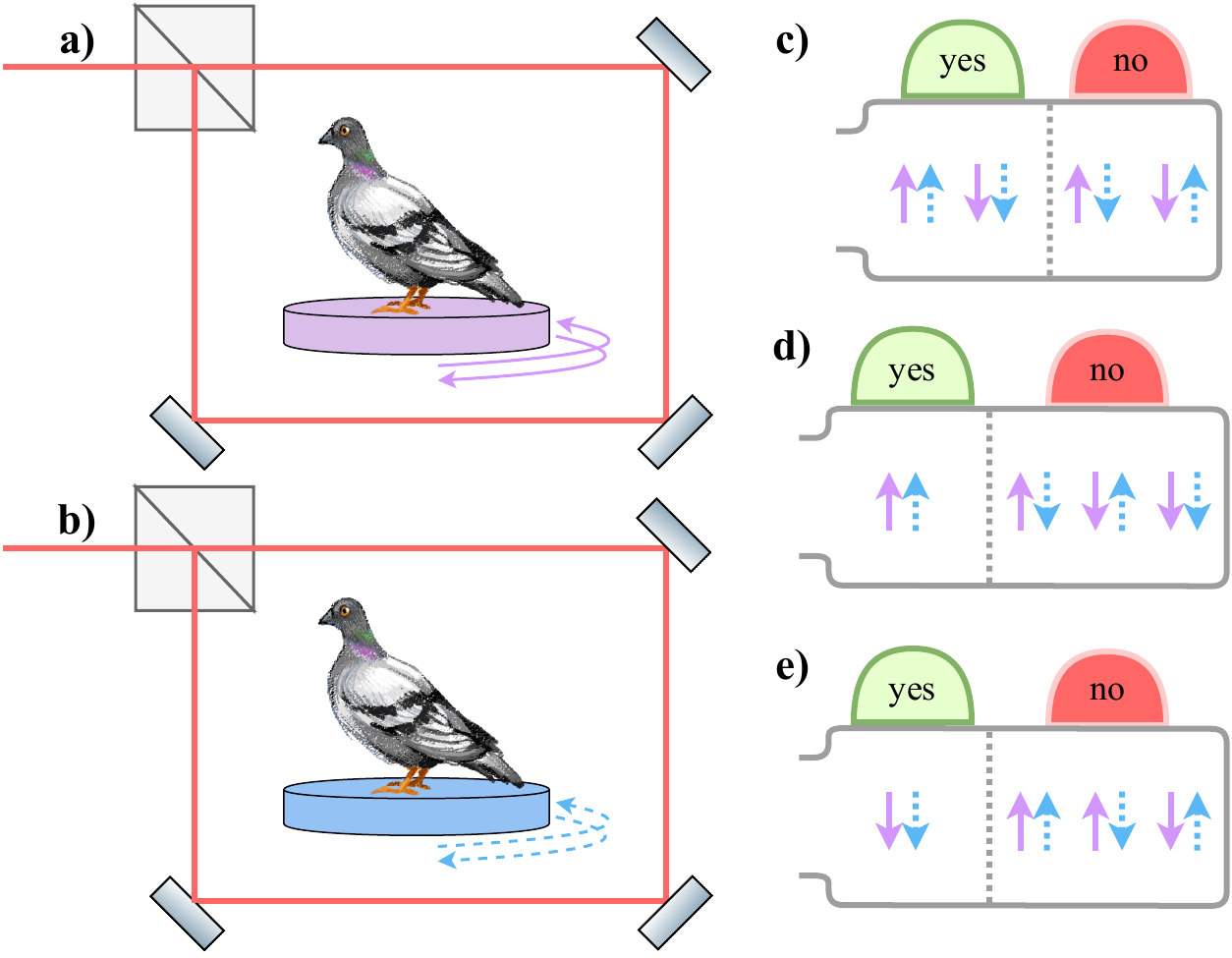}
    \caption{Conceptual overview. a) and b): quantum pigeons in two pigeonholes are represented as photons that can propagate clockwise or anti-clockwise through a displaced Sagnac interferometer. Beamsplitters prepare the pigeons in $\ket{+}\ket{+}$, then one of the devices c), d), or e) measures them, then we postselect them onto the state $\ket{+i}\ket{+i}$ by only detecting photons that exit the same port of the interferometer they entered.  c), d), and e): different measurements used to observe the pigeons. c) asks ``are both pigeons traveling the same direction?'' d) asks ``are both pigeons traveling clockwise?'' e) asks ``are both pigeons traveling anti-clockwise?''.}
    \label{fig:pigeonhole}
\end{figure}

In our experiment, the ``pigeons'' are photons in displaced Sagnac interferometers and the ``pigeonholes'' are the two possible travel directions: clockwise $\ket{\L}$ and anti-clockwise $\ket{\R}$  (see Fig. \ref{fig:pigeonhole}). 
The corresponding, observables for a particular pigeon $k \in \{ (1, 2, 3) \}$ are the projectors $\PL^k=\ket{\L}^k\bra{\L}^k$ and $\PR^k=\ket{\R}^k\bra{\R}^k$. Each photon is preselected in the state $\ket{+}^k = \left (\frac{\ket{\L}^k + \ket{\R}^k}{\sqrt{2}} \right )$ as it enters the interferometer and later postselected in the state  $\ket{+i}^k = \left (\frac{\ket{\L}^k + i\ket{\R}^k}{\sqrt{2}} \right )$ as it leaves. The quantum pigeonhole paradox concerns the observables $\PS{k, \ell}=\PR^k \PR^\ell + \PL^k \PL^\ell$, which asks whether a particular pair of photons $(k, \ell) \in \{ (1, 2), (1, 3), (2, 3) \}$ traveled the same direction.

These observables seems to defy two fundamental counting principles.
\begin{enumerate}
    \item \textbf{The pigeonhole principle}: when there are more pigeons than holes, at least one pair must be in the same hole. Violated because $|\bra{+i}^k \bra{+i}^\ell \PS{k, \ell} \ket{+}^k \ket{+}^\ell|^2 = 0$.
    \item \textbf{The sum rule}: the number of pigeons among two holes is at least the number of pigeons in either hole. Violated because $|\bra{+i}^k \bra{+i}^\ell \PL^k \PL^\ell \ket{+}^k \ket{+}^\ell|^2 = 1/16 > 0$ and $|\bra{+i}^k \bra{+i}^\ell \PR^k \PR^\ell \ket{+}^k \ket{+}^\ell|^2 = 1/16 > 0$.
\end{enumerate}

\section*{The quantum pigeonhole paradox}

As a quantum paradox, the pigeonhole paradox exhibits several crucial features. First, the paradox involves correlations in a non-local system: each pigeon could be in a different space-like separated region of the galaxy, but quantum mechanics predicts the paradox will occur nevertheless. Second, the preselection and postselection are both separable states, making it difficult to rely on the features of quantum entanglement to explain away the paradox. Finally, these separable states are symmetric, making it irrelevant which particular pair of pigeons are observed.

This last feature is experimentally convenient, as it enables us to measure only a single pair of pigeons. While putting two pigeons in two different holes is trivial, we can still derive a bound that separates intuitive classical behavior from the phenomna arising in the pigeonhole paradox. Suppose three pigeons are placed in two holes according to an arbitrary probability distribution $P(b_1, b_2, b_3)$, where $b_i \in \{ \L, \R \}$ are boolean variables indicating which of two holes the $i$th pigeon is placed. Uniformly at random, a pair of pigeons are examined to determine whether or not they are in the same hole. 
The chance to find the random pair in the same hole is
\begin{align*}
    ( & P(\L, \L, \L) + P(\R, \R, \L) + P(\L, \L, \R) + P(\R, \R, \R) \\
    + & P(\L, \L, \L) + P(\R, \L, \R) + P(\L, \R, \L) + P(\R, \R, \R) \\
    + & P(\L, \L, \L) + P(\L, \R, \R) + P(\R, \L, \L) + P(\R, \R, \R))/3 \\
    = 1&/3  +(P(\L, \L, \L) + P(\R, \R, \R))  2/3 \\
    \geq 1&/3.
\end{align*}
There is no classical distribution of three pigeons across two holes in which a random pair of pigeons is found in the same hole less than $1/3$ of the time. In this sense, a measurement of $\PS{\ell k}$ that comes out ``yes'' with any frequency less than $1/3$ violates the classical pigeonhole principle. For the rest of the manuscript, we will deal only with Pigeons $1$ and $2$  without loss of generality, so we set $\PS{} = \PS{12}$, $\PLL{} = \PL^1\PL^2$, and $\PRR{} = \PR^1 \PR^2$.

Our experiment hones in on the distinction between an observable and the variety of ways to measure that observable. We compare several different ways to measure $\PS{}$. The first has only two outcomes: \{``same direction'',  ``different direction''\}. That is, the measurement reveals whether photon 1 and 2 travelled the same direction, but not which direction they travelled. The second sums the results of \{``both clockwise'', ``neither clockwise'' \} and \{``both anti-clockwise'', ``neither anti-clockwise'' \}. 

\begin{figure}
    \centering
    \includegraphics[width=\columnwidth]{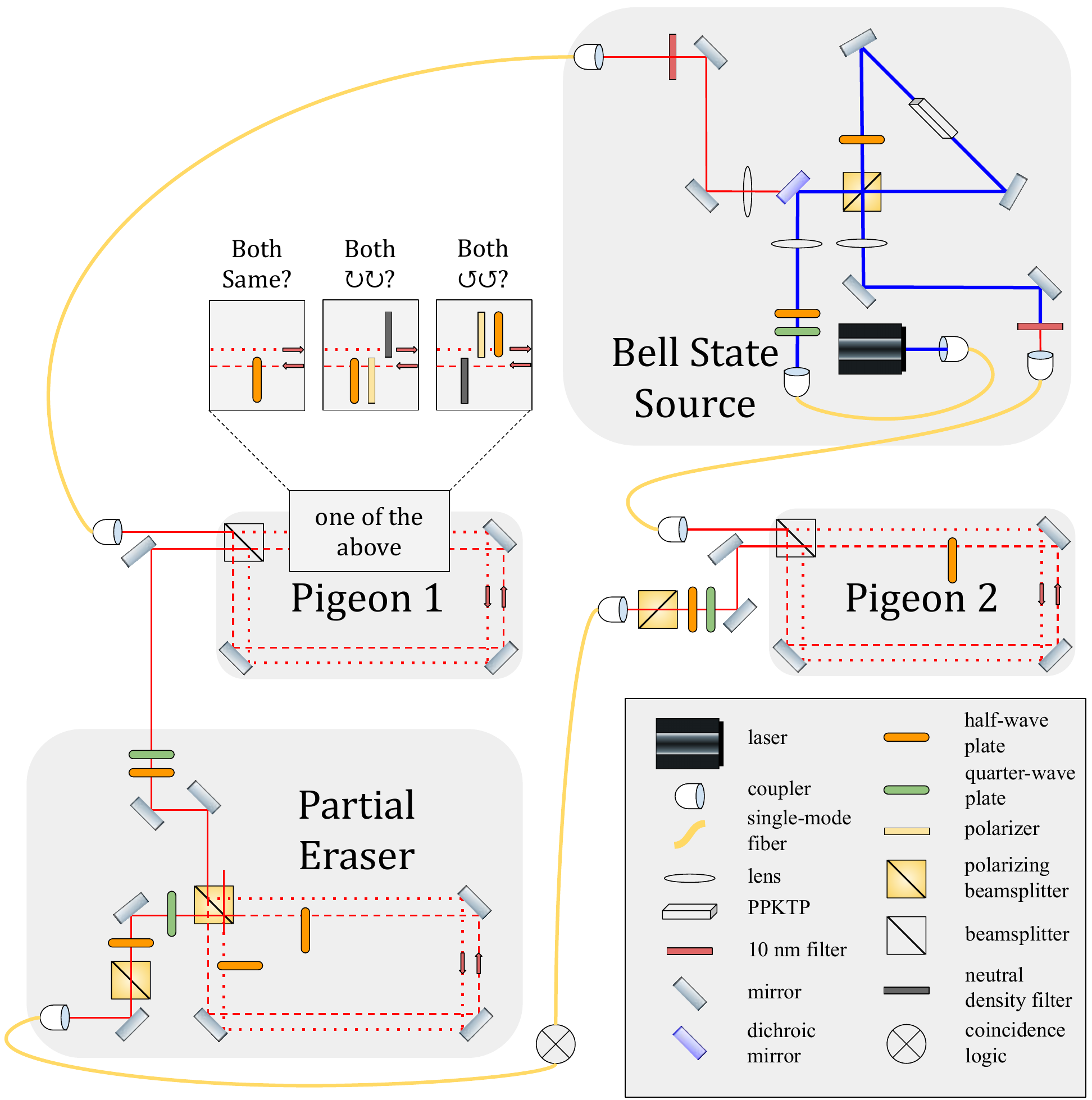}
    \caption{Experimental setup. A Bell-state source generates polarization-entangled photons by pumping a PPKTP crystal with a continuous-wave laser at $405$ nm. Each photon enters a non-polarizing displaced Sagnac interferometer which allows the photonic pigeon to travel in a superposition of two pigeonholes. The photons are postselected by ignoring an output port of each interferometer. A half waveplate in the anti-clockwise path of the Pigeon 2 interferometer performs a strong which-path measurement. The photon is detected and the result of its which-path measurement steers Pigeon 1. One of three different sets of optics in Pigeon 1 encode a strong, direct measurement of either $\PS{}$, $\PLL{}$, or $\PRR{}$ onto polarization. The partial eraser, a  displaced Sagnac interferometer, couples polarization to path, then erases the polarization information. The strength of the polarization-path coupling is set by the angle of two half waveplates inside the partial eraser.  }
    \label{fig:experiment}
\end{figure}

Every ``yes'' or ``no'' question in quantum mechanics can be represented by some projector $\Pi$. We will call a measurement of $\Pi$ ``direct'' if it indicates whether or not the system was in a ``yes'' (or ``no'') state, but not which among potentially many ``yes'' (or ``no'') states the system was in. Mathematically, we represent such a ``direct'' measurement by a unitary $U_\Pi(s)$ that acts on any system state $\ket{\psi}$ and a particular meter state $\ket{\mu}$ according to
\begin{equation}
\label{eq:unitary_coupling}
    U_\Pi(s) \ket{\psi}\ket{\mu} = \Pi \ket{\psi} \ket{s} + (\id - \Pi) \ket{\psi} \ket{-s},
\end{equation}
where
\begin{equation}\label{eq:measurement_strength}
    s = \sqrt{1 - |\braket{s | -s}|^2}
\end{equation}
is the ``strength'' of the measurement. A strong measurement has $s = 1$ and sends the meter into one of two orthogonal state: $\ket{s = 1}$ or $\ket{s = -1}$. A weak measurement has $s \ll 1$, meaning the meter shifts to one of two nearly indistinguishable states.

A single weak measurement provides vanishing information about whether the system was in a ``yes'' or ``no'' state, but averaging the results of weak measurements across many independent and identical systems yields a meaningful aggregate, the so-called ``weak value'' \cite{AAV}. 
The weak value of an observable $\Pi$ prepared in state $\ket{\psii}$ and postselected in state $\ket{\psif}$ is the complex number
\begin{equation} \label{eq:weak_value}
    \frac{\bra{\psif} \Pi \ket{\psii}}{\braket{\psif | \psii}}.
\end{equation}
The real part of the weak value is obtained by measuring the meter in the ``real'' basis $\ket{\pm \Real} = \ket{s = \pm 1}$. The imaginary part comes from measuring the meter in the ``imaginary'' basis $\ket{\pm \Imag} = (\ket{+ \Real} \pm i \ket{-\Real}) / \sqrt{2}$.

In the strong limit, measuring the meter in the real basis gives the probability for the system to have been in a ``yes'' state, given the postselection succeeded. This probability is predicted by the Aharonov-Bergmann-Lebowitz formula \cite{ABL}.
\begin{equation}\label{eq:ABL_rule}
    \frac{|\bra{\psif} \Pi \ket{\psii}|^2}{|\bra{\psif} \Pi \ket{\psii}|^2 + |\bra{\psif} (\id - \Pi) \ket{\psii}|^2}
\end{equation}
Curiously, the literature on PPS experiments has so far ignored the strong limit of measuring the meter in the imaginary basis. Even the recent work of De Zela, which studied the role of weak values in strong measurements, did not comment on such strong imaginary measurements\cite{de2022role}. Like their weak counterparts \cite{Dressel12}, they can be related to the sensitivity of the postselection probability to the measurement's back-action (see Sec. \ref{sec:imaginary} for more details). The usual model of quantum measurement uses a Gaussian probe, and in fact, this model predicts the imaginary part goes to $0$ in the infinitely strong limit \cite{Dressel12}. However, the decay of the imaginary part to $0$ is not a universal feature of quantum measurement. For a strong, direct measurement of a projector via a qubit probe, the quantity comes out to
\begin{equation}
    \frac{\Imag[\braket{\psii | \psif}\bra{\psif} \Pi \ket{\psii}]}{|\bra{\psif} \Pi \ket{\psii}|^2 + |\bra{\psif} (\id - \Pi) \ket{\psii}|^2}.
\end{equation}
This value is not always $0$, which complicates criticisms of the weak value on the basis of its imaginary component \cite{Svensson13, Svensson_2014, KASTNER200457}. Strong measurements, at least in some sense, have imaginary parts too.

\section*{Direct non-local measurements}

Our experiment implements three direct measurements at a continuum of measurement strengths: $U_\PS{}(s)$, $U_\PLL{}(s)$, and $U_\PRR{}(s)$. These measurements are made on pairs of particles, and must be non-destructive \cite{Paraoanu2018}.  These  types of non-local measurements  are currently attracting considerable  attention with a number of recent theoretical proposals \cite{BCnonlocal,BCsequential,ReschSteinberg,BrodutchVaidman,Wu2016,Kedem2010,edamatsu2016,yokota2018} and  experimental results at both the strong limit \cite{Xu2019,Li2019,Pan2019} and  the weak limit \cite{Lundeen2009,Yokota2009,PhysRevLett.123.150402}. The Hamiltonians that generates variable-strength, direct measurements of two-body observables such as $\PLL{}$ require three-body interaction terms of the form
\begin{equation}
    M^0 \Pi^1 \Pi^2,
\end{equation}
where $M^0$ is some meter observable. Physically, these terms represent a single meter interacting simultaneously with two spatially separate systems. This interaction would have to be non-local, and thus un-physical. Nevertheless, it is possible to simulate these interactions using a combination of quantum steering and quantum erasure \cite{BCnonlocal}.

The first step in generating these non-local interactions is to prepare our photonic pigeons with a shared and entangled polarization probe. We use the Bell state source, illustrated in Fig. \ref{fig:experiment} and described in detail in Sec. \ref{sec:source} to generate pairs of photons whose joint polarization state is $\frac{\ket{H}\ket{H} + \ket{V}\ket{V}}{\sqrt{2}}$. We use the standard notation $\ket{H}$, $\ket{V}$ for horizontal and vertical polarization respectively, $\ket{D} = (\ket{H} + \ket{V})/\sqrt{2}$, $\ket{A} = (\ket{H} - \ket{V})/\sqrt{2}$ for diagonal and anti-diagonal respectively, and $\ket{R}=(\ket{H} +i  \ket{V})/\sqrt{2}$, $\ket{L}=(\ket{H} -i  \ket{V})/\sqrt{2}$ for right- and left-circular polarizations, respectively.  Each photon is then sent into its own displaced Sagnac interferometer, which is opened and closed with a 50/50 beamsplitter to create the initial and final states. The initial state is always  $\ket{\psii}=\ket{+}\ket{+}$ and the final state, which is selected by only detecting photons that exit the interferometer through the same port they entered and optionally blocking one of the paths is $\ket{\psif} \in \{\ket{+i}\ket{+i}$, $\ket{\L}\ket{\L}, \ket{\L}\ket{\R}, \ket{\R}\ket{\L}, \ket{\R}\ket{\R}\}$.

Polarization optics in the Sagnc interferometers perform strong, but non-destructive which-path measurements. The Pigeon 2 interferometer has a half waveplate in its anti-clockwise path, but not its clockwise path. Projecting photon 2 onto $\ket{D}$ (which succeeds with probability $1/2$) steers Photon 1's initial polarization to $\ket{D}$ if photon 2 went clockwise and $\ket{A}$ if photon 2 went anti-clockwise. At this stage, the polarization-path state (with normalization indicating success probability) is
\begin{equation}
    \left (\ket{D}^1 \otimes \PL^2 \ket{\psii}^{1 2} + \ket{A}^1 \otimes \PR^2 \ket{\psii}^{1 2} \right ) / \sqrt{2}.
\end{equation}

One of three different strong which-path measurements (shown in Fig. \ref{fig:experiment}), corresponding to $\PS{}$, $\PLL{}$, and $\PRR{}$, shifts the polarization of photon 1. To measure $\PS{}$, we place a half waveplate in the anti-clockwise path, but not in the clockwise path, which evolves the state to
\begin{equation}
    \left (\ket{D}^1 \otimes \PS{1 2} \ket{\psii}^{1 2} + \ket{A}^1 \otimes (\id - \PS{1 2}) \ket{\psii}^{1 2} \right ) / \sqrt{2}.
\end{equation}
The polarization of photon 1 at this stage strongly encodes whether both photons have traveled the same direction, but not which direction they traveled.

To measure $\PLL{}{}$, we place a polarizer followed by a half waveplate in the anti-clockwise path that together apply the operator $-\ket{A}\bra{V}$. Photon 1 only survives this operator with probability $1/2$. To keep the interferometer balanced, we place a neutral density filter with transmission probability $1/2$ in the clockwise path. The state at this point is
\begin{equation}
    \left (\ket{D}^1 \otimes \PLL{1 2} \ket{\psii}^{1 2} + \ket{A}^1 \otimes (\id - \PLL{1 2}) \ket{\psii}^{1 2} \right ) / \sqrt{4}.
\end{equation}

The approach for $\PRR{1 2}$ is similar. In the clockwise path, we place a polarizer and half waveplate to effect the operator $-\ket{D}\bra{V}$ and in the other path we place a neutral density filter. The state becomes
\begin{equation}
    \left (\ket{D}^1 \otimes (\id - \PRR{1 2}) \ket{\psii}^{1 2} + \ket{A}^1 \otimes \PRR{1 2} \ket{\psii}^{1 2} \right ) / \sqrt{4}.
\end{equation}

The next step for all three measurements is to project the path state onto $\ket{\psif}$ by closing the Pigeon 1 interferometer. A quarter and half waveplate apply a unitary $U$, which rotates the results of the strong measurement into either the $H/V$ basis, for a ``real'' measurement or the $R/L$ basis for an ``imaginary'' measurement. The state of Photon 1's polarization at this point is proportional to
\begin{equation}
    \bra{\psif} \Pi \ket{\psii} U \ket{D} + \bra{\psif}(\id-\Pi)\ket{\psii} U \ket{A}
\end{equation}
for any of the three projectors $\Pi$.

\begin{figure*}
    \centering
    \includegraphics[width=1.9\columnwidth]{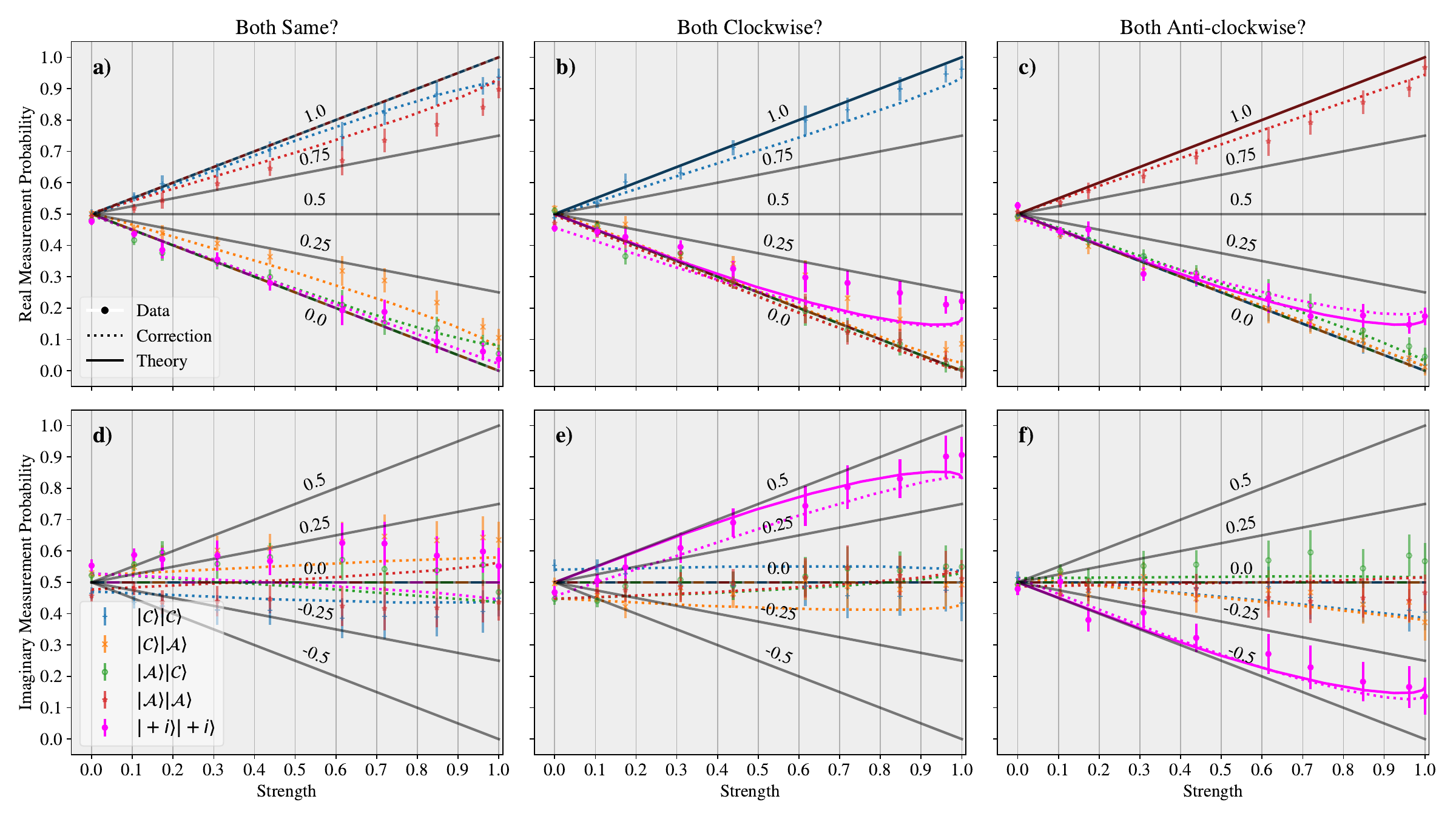}
    \caption{Variable-strength measurement results. In each plot, the $x$-axis denotes measurement strength (Eq. \ref{eq:measurement_strength}) such that $0$ is a weak measurement and $1$ is a strong measurement. a), b), and c) show the results for direct measurements of the projectors $\PS{12}$, $\PLL{12}$, and $\PRR{12}$ respectively. The $y$-axis is the probability for the polarization meter to be in the real ``yes'' state. The labeled grid slopes denote the corresponding probability for the pigeons to be in the ``yes'' subspace of the projector, after accounting for measurement strength. At small measurement strengths, these values correspond to the real part of the weak value. Each marker and color combination represents a different path postselection. The fuchsia dots represent the paradoxical postselection $\ket{+i}\ket{+i}$. The other four colors are calibration data. Two different theoretical models are plotted for comparison. Solid lines are theory. Dotted lines represent a correction to this theory tomographic calibrations. d), e), and f) are organized similarly, but their $y$-axes show the results of measuring in the imaginary basis. At small measurement strengths, the values denoted by the labelled grid slopes correspond to the imaginary part of the weak value. The direct measurements of $\PS{12}$ suggest that photons postselected in $\ket{+i}\ket{+i}$ propagate in different directions at least as often as photons which always take the $\ket{\L}\ket{\R}$ path or the $\ket{\R}\ket{\L}$ path at all strengths. The disturbance to the meter along its imaginary axis for photons postselected in $\ket{+i}\ket{+i}$ is somewhere between that of $\ket{\L}\ket{\R}$ path or the $\ket{\R}\ket{\L}$ for all strengths. Thus, measurement disturbance cannot explain why $3$ pigeons among $2$ pigeonholes can each be in a different hole. On the other hand, direct measurements of $\PLL{12}$ and $\PRR{12}$ show that measurement disturbance is responsible for violation of the sum rule. Strong measurements suggest photons postselected in $\ket{+i}\ket{+i}$ both travel clockwise/anti-clockwise a significant fraction of the time, but at weaker measurements, they fall within the ``no'' calibration points and agree with the direct measurement of $\PS{12}$.}
    \label{fig:data}
\end{figure*}

We use a quantum eraser to erase some, but not all, of the coupling created by the strong measurement. We send the photon into a partial eraser (see Fig. \ref{fig:experiment}), which we implement as a displaced Sagnac interferometer that opens and closes with a polarizing beamsplitter. The interferometer couples polarization and path with a strength determined by the angle of two waveplates that enact equal and opposite rotations to photons traveling through either of the two interferometer paths. The strong correlations between polarization and the original pigeon observable are destroyed by projecting the polarization onto an unbiased basis, which incurs another loss factor of $1/2$. The path state of the photon just before being detected is proportional to
\begin{equation}
    \bra{\psif} \Pi \ket{\psii} \ket{s} + \bra{\psif}(\id-\Pi)\ket{\psii} \ket{-s},
\end{equation}
where
\begin{equation}
    \ket{s} = \sqrt{\frac{1 + s}{2}} \ket{\mathrm{yes}} + \sqrt{\frac{1 - s}{2}} \ket{\mathrm{no}}.
\end{equation}
In a ``real'' measurement, $\ket{\mathrm{yes}}$ and $\ket{\mathrm{no}}$ correspond the output ports of the partial eraser. The imaginary measurement rotates these states to the superpositions $(\ket{\mathrm{yes}} \pm i \ket{\mathrm{no}}) / \sqrt{2}$. All told, the variable-strength non-local measurement procedure works with probability $1/4$ for $\PS{}$ and probability $1/8$ for $\PLL{}$ or $\PRR{}$.

\section*{Results}

The results of our experiment are plotted in Fig. \ref{fig:data}. Our goal is to observe properties of the pigeon system, but our data come from measurements of a polarization meter. We call the probability of finding our polarization probe in its real ``yes'' state the ``real meter value'' and the probability for the pigeon system to be in a ``yes'' state of our pigeon observable the ``real system value''. In a strong measurement, the real meter value is simply equal to the real system value. For weaker measurements, pigeons in a ``yes'' state of an observable only shift the meter by a small amount, leaving the meter in the real ``no'' state some fraction of the time. To convert meter values to system values, we divide by measurement strength.

\begin{figure*}
    \centering
    \includegraphics[width=1.9\columnwidth]{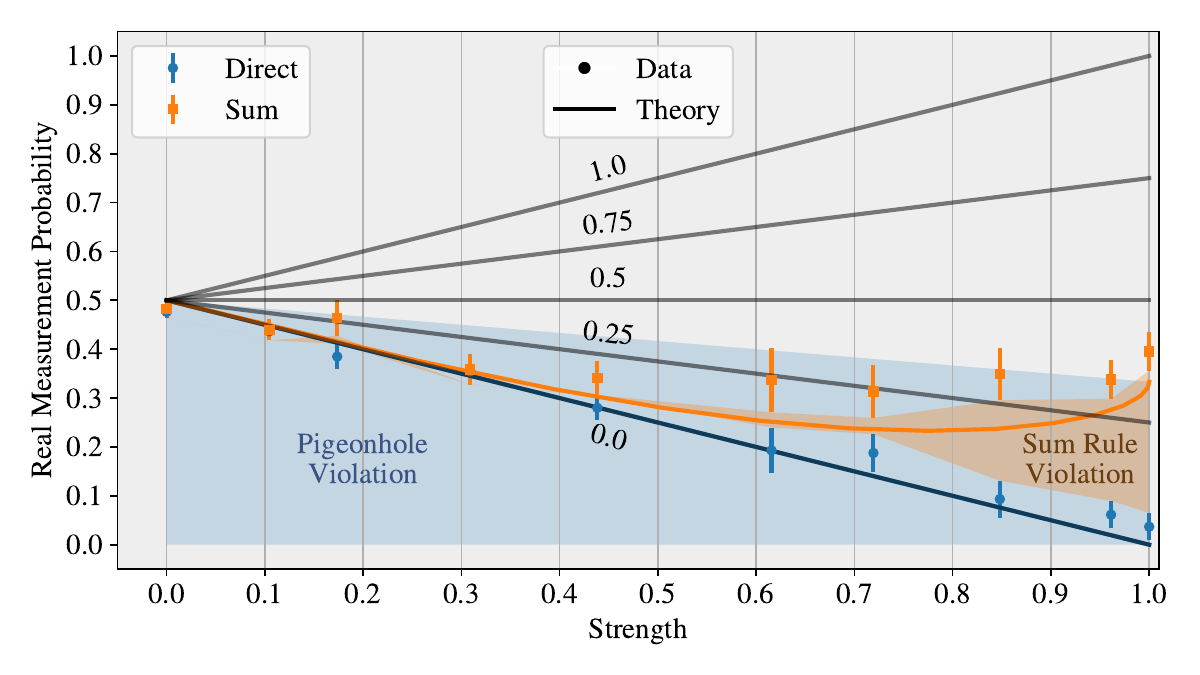}
    \caption{Violation of classical counting principles. The $x$-axis denotes measurement strength (Eq. \ref{eq:measurement_strength}) such that $0$ is a weak measurement and $1$ is a strong measurement. The $y$-axis is the probability for the polarization meter to be in the real ``yes'' state. The labeled grid slopes denote the corresponding probability for the pigeons to be in the same hole after accounting for measurement strength. Blue dots show results from direct measurements of the same hole projector $\PS{}$. Orange squares denote an indirect measurement obtained from summing the direct measurements for $\PLL{}$ and $\PRR{}$. Blue and orange slopes denote theoretical predictions for the respective points. The orange region highlights the difference between our two methods for measuring $\PS{}$, indicating a violation of the sum rule. The violation shrinks as the measurement strength decreases. The blue region indicates where direct measurements of $\PS{}$ would violate the pigeonhole principle. At all strengths, our direct measurements of $\PS{}$ fall in this region.}
    \label{fig:violation}
\end{figure*}

Similarly, we call the probability of finding our polarization probe in the imaginary ``yes'' state the ``imaginary meter value''. To convert to the corresponding ``imaginary system value'', we subtract $1/2$ (so that a probe in the real ``yes'' state registers as having $0$ imaginary component) and then divide by measurement strength. In the weak limit, the real and imaginary system value equal the real and imaginary component of the weak value (Eq. \ref{eq:weak_value}). In the strong limit, the real system value equals the ABL probability (Eq. \ref{eq:ABL_rule}). The strong limit of the imaginary system value has not been described previously in the literature, but it is related to the sensitivity of the postselection probability to measurement back-action (see Sec. \ref{sec:imaginary} for more details).

All data in Fig. \ref{fig:data} are plotted according to their meter value. Grid slopes represent the conversion between meter values and system values. We begin with the calibration points, which were taken by blocking one path in each pigeon interferometer to postselect the state on one of $\ket{\L}\ket{\L}, \ket{\L}\ket{\R}, \ket{\R}\ket{\L},$ or $\ket{\R}\ket{\R}$. These photons should all have a real system value of either $0$ or $1$ and their real meter values should move towards $1/2$ linearly in the measurement strength. Their imaginary meter value should be $1/2$ and their imaginary system value should be $0$ at all measurement strengths. These predictions (solid lines in Fig. \ref{fig:data}) agree qualitatively with our data (points in Fig. \ref{fig:data}). 

The data postselected on $\ket{+i}\ket{+i}$ are the meat of the experiment (fuchsia dots in \ref{fig:data}). The direct measurements of $\PS{}$ (\ref{fig:data} a) reveal that the real system value for both pigeons to be in the same hole is indeed consistent with $0$ at all measurement strengths. This can be seen from the fact that the real meter value of the $\ket{+i}\ket{+i}$ data is never significantly higher than the $\ket{\L}\ket{\R}$ or $\ket{\R}\ket{\L}$ calibration data. While the imaginary values of the $\ket{+i}\ket{+i}$ data do not quite match their predicted value of $0$, they are no further from this prediction than the calibration data. Thus, measurement disturbance is not a viable resolution to the quantum pigeonhole paradox.

On the other hand, our data confirm that measurement disturbance does explain the violation of the sum rule. The real part of our $\ket{+i}\ket{+i}$ data is ``both clockwise'' or ``both anti-clockwise'' significantly more often than it is ``both same'' at high strengths, but agrees with the ``both same'' values at weaker strengths. For both strong and weak measurements, the imaginary part of ``both same'' significantly differs from the imaginary parts of ``both clockwise'' and ``both anti-clockwise''. Nevertheless, the sum rule holds because the imaginary parts of ``both clockwise'' and ``both anti-clockwise'' are equal and opposite. While the cancellation of the imaginary parts in the weak limit was predicted by the weak value formula \ref{eq:weak_value}, it is remarkable that the cancellation extends into the strong regime, even as the sum rule for the real part fails. The question of whether this kind of cancellation is a general feature of imaginary system values at all measurement strengths is a fascinating avenue for further study.

Having explained the qualitative features of our data, we turn to its quantitative agreement with theory. Most of the discrepancy between our data and theory is accounted for by the quality of our Bell state source and polarization optics. The fidelity of our actual Bell state to the desired one is $95\%$. Furthermore, the nominally non-polarizing beamsplitters used to open and close our pigeon interferometers are birefringent, complicating our polarization-based which-path measurements. We measure these effects using polarization state tomography of our Bell state source and polarization process tomography in each path of our pigeon interferometer and use them to generate refined predictions, plotted as dotted lines in Fig. \ref{fig:data}. The systematic uncertainty of our data dominates its statistical uncertainty. The error bars in Fig. \ref{fig:data} and Fig. \ref{fig:violation} denote the root-mean-square deviation of our calibration data from these refined predictions. They are computed separately for each measurement strength and quadrature (real or imaginary).

To conclude, we realized the first variable-strength measurement of a non-local observable to study quantum violations of two seemingly irrefutable counting laws: the pigeonhole principle and the sum rule. Our data show that the violation of the sum rule is an artifact of measurement disturbance. On the other hand, the violation of the pigeonhole principle is the same, regardless of the strength of the back-action. Our experiment shows there is a concrete, empirical, and measurement-independent sense in which three quantum pigeons really can occupy two holes without any being in the same hole. Finally, our variable-strength apparatus led us to discover that the same process that yields the imaginary part of a weak measurement can yield a non-trivial quantity in the strong limit as well. Further exploration of the quantity is a ripe opportunity for further research.

\section*{Supporting Information Appendix (SI)}

\subsection{Variable-strength measurements with a qubit meter}

The direct measurements of projectors $\Pi_\S$ on the pigeon system are described by the von Neumann interaction unitary
\begin{equation}
    U_{\S\Ob} = e^{-i \theta (2 \Pi_\S - \id_\S) \sigma_{y, \Ob}}
\end{equation}
between the pigeon system $\S$ and an observer $\Ob$ with a spin-$1/2$ meter.
The meter momentum $\sigma_{y, \Ob}$ generates rotations of the initial meter state $\ket{+z}_\Ob$ by $\pm \theta$ radians in Hilbert space depending on whether the initial pigeon state $\ket{\psii}_\S$ is in a ``yes'' (eigenvalue 1) or ``no'' (eigenvalue 0) eigenspace of $\Pi_\S$. Euler's theorem simplifies the unitary to
\begin{align}
    U_{\S\Ob} & = \Pi_\S (\cos \theta \id_\Ob -i \sin \theta \sigma_{y, \Ob}) \\
    & + (\id - \Pi_\S) (\cos \theta \id_\Ob + i \sin \theta  \sigma_{y, \Ob}).
\end{align}
This unitary evolves the initial system and meter state to
\begin{equation}
    U_{\S\Ob} \ket{\psii}_\S \ket{+z}_\Ob = \Pi_\S \ket{\psii}_\S \ket{s}_\Ob + (\id - \Pi_\S) \ket{\psii}_\S \ket{-s}_\Ob,
\end{equation}
where
\begin{equation}
    \ket{s}_\Ob = \sqrt{\frac{1 + s}{2}} \ket{+x}_\Ob + \sqrt{\frac{1 - s}{2}} \ket{-x}_\Ob,
\end{equation}
and $s = \sin \theta$. The meter observable $\Pi_\Ob$ for the real basis is calibrated so that
\begin{equation}
    \braket{\pm s | \Pi_\Ob | \pm s} = \frac{1 \pm 1}{2},
\end{equation}
which is satisfied when
\begin{equation}
    \Pi_\Ob = \frac{\id + \sigma_{x, \Ob} / s}{2}.
\end{equation}
The real part of the system value is the expectation of $\Pi_\Ob$ conditioned on successful postselection
\begin{equation}
    \E \left [ \Pi_\Ob | \ketbra{\psif}{\psif}_\S \right ] = \frac{\E \left [ \ketbra{\psif}{\psif}_\S \Pi_\Ob \right ]}{\E \left [ \ketbra{\psif}{\psif}_\S \right ]},
\end{equation}
where the expectation value is taken over the coupled state $U_{\S\Ob} \ket{\psii}_\S \ket{+z}_\Ob$.
The denominator is the probability that the postselection succeeds.
\begin{align}
    \E \left [ \ketbra{\psif}{\psif}_\S \right ] 
    & = |\braket{\psif | \Pi_\S | \psii}|^2 + |\braket{\psif | (\id - \Pi_\S) | \psii}|^2 \\
    & + 2 \sqrt{1 - s^2} \Real[\braket{\psif | \Pi_\S | \psii}\braket{\psii | (\id - \Pi_\S) | \psif}].
\end{align}
The numerator is the joint expectation
\begin{align}
    \E \left [ \ketbra{\psif}{\psif}_\S \Pi_\Ob \right ] 
    = & |\braket{\psif | \Pi_\S | \psii}|^2 + \sqrt{1 - s^2} \times \\ 
    & \Real[\braket{\psif | \Pi_\S | \psii}\braket{\psii | (\id - \Pi_\S) | \psif}].
\end{align}
When $s = 0$, the conditional expectation simplifies to the real part of the weak value (Eq. \ref{eq:weak_value}) and when $s = 1$, it simplifies to the Aharonov-Bermann-Lebowitz formula \cite{ABL} (Eq. \ref{eq:ABL_rule}).

\subsection{Imaginary measurements}\label{sec:imaginary}

To measure the imaginary part of the weak value, we measure in the imaginary meter basis $\sigma_{y, \S}$. The imaginary meter observable $\Tilde{\Pi}_\Ob$ is calibrated so that
\begin{equation}
    \braket{\pm s | \Tilde{\Pi}_\Ob | \mp s} = \pm i / 2,
\end{equation}
which is satisfied with
\begin{equation}
    \Tilde{\Pi}_\Ob = \frac{\sigma_{y, \Ob}}{2 s}.
\end{equation}
The conditional expectation of the imaginary meter observable is
\begin{equation}
    \E \left [ \Tilde{\Pi}_\Ob | \ketbra{\psif}{\psif}_\S \right ] = \frac{\E \left [ \ketbra{\psif}{\psif}_\S \Tilde{\Pi}_\Ob \right ]}{\E \left [ \ketbra{\psif}{\psif}_\S \right ]}.
\end{equation}
The denominator is the postselection probability probability from the real case treated earlier. The numerator is
\begin{equation}
    \E \left [ \ketbra{\psif}{\psif}_\S \Tilde{\Pi}_\Ob \right ] = \Imag[\braket{\psif | \Pi_\S | \psii}\braket{\psii | (\id - \Pi_\S) | \psif}].
\end{equation}
When $s = 0$, the conditional expectation simplifies to the imaginary part of the weak value (Eq. \ref{eq:weak_value}). Curiously, the joint expectation $\E \left [ \ketbra{\psif}{\psif}_\S \Tilde{\Pi}_\Ob \right ]$ does not depend on the measurement strength $s$ at all. If the imaginary part of the weak value is non-zero, the conditional expectation of the imaginary meter observable $\Tilde{\Pi_\Ob}$ will be non-zero at all measurement strengths. However, this result is specific to a spin-$1/2$ meter. If the meter was instead the position of a particle, the Euler expansion of the coupling unitary used to evaluate the evolution at all measurement strengths would not be valid.

We have shown how to calculate the conditional expectation of the imaginary meter observable, but we have so far said nothing on how to interpret it. The imaginary part of the weak value is related to the sensitivity of the postselection success probability to the back-action the meter suffers from measuring a system observable \cite{Dressel12}. The probability $P_\ps$ of successfully postselecting an initial system state $\ket{\psii}_\S$ onto a final system state $\ket{\psif}_\S$ is
\begin{align}
    P_\ps 
    & = ||\bra{\psif}_\S e^{-i \theta (2 \Pi_\S - \id_\S) \sigma_{y, \Ob}} \ket{\psii}_\S \ket{+z}_\Ob||^2. \\
    & = \E \left [ e^{i 2 \theta \ad_{\Pi_\S} \sigma_{y, \Ob}} \left [ \ketbra{\psif}{\psif}_\S \right ] \right ].
\end{align}
The super operator $\ad_{\Pi_\S}$ represents the infinitesimal back-action due to measuring $\Pi_\S$. 
The action of $\ad_{\Pi_\S}$ on an arbitrary system operator $A_\S$ is $\ad_{\Pi_\S} [A_\S] = \Pi_\S A_\S - A_\S \Pi_\S$. 
To describe how the postselection probability changes with the strength of the back-action, we will increase the size of $\ad_{\Pi_\S}$ uniformly with the transformation 
$i \theta \ad_{\Pi_\S} \rightarrow i \theta \ad_{\Pi_\S} + \delta \id_\S / \sin(\theta)$. $\delta$ is an artificial parameter that lets us tune the strength of the back-action independently from the measurement strength. The postselection probability $P_\ps(\delta)$ as a function of this back-action parameter is
\begin{equation}
    P_\ps(\delta) = \E \left [ e^{i 2 \theta \ad_{\Pi_\S} \sigma_{y, \Ob} + 2 \delta \Tilde{\Pi}_\Ob} \left [ \ketbra{\psif}{\psif}_\S \right ] \right  ]
\end{equation}
using the fact that $\Tilde{\Pi}_\Ob = \sigma_{y, \Ob} / (2 \sin \theta)$.
The sensitivity of the postselection probability to back-action is
\begin{equation}
    \frac{\partial \log P_\ps}{\partial \delta} \biggr\rvert_{\delta = 0} = 2 \frac{\E \left [ e^{i 2 \theta \ad_{\Pi_\S} \sigma_{y, \Ob}} \left [ \ketbra{\psif}{\psif}_\S \Tilde{\Pi}_\Ob \right ] \right ]}{\E \left [ e^{i 2 \theta \ad_{\Pi_\S} \sigma_{y, \Ob}} \left [ \ketbra{\psif}{\psif}_\S \right ] \right ]}.
\end{equation}
The right-hand side of this expression is exactly twice the conditional expectation of the imaginary meter observable.
\begin{equation}
    \frac{1}{2} \frac{\partial \log P_\ps}{\partial \delta} \biggr\rvert_{\delta = 0} = \E \left [ \Tilde{\Pi}_\Ob | \ketbra{\psif}{\psif}_\S \right ]
\end{equation}
In the weak limit, this conditional expectation equals the imaginary part of the weak value. In the strong limit, it may not manifestly be the imaginary part of a complex number, but it is still related to the sensitivity of the postselection probability in the same way.

\subsection{Entangled photon source}\label{sec:source}

We create photon pairs with a wavelength near $810$ nm via type II colinear spontaneous parametric down-conversion. We pump a periodically polled potassium titanyl phosphate (PPKTP) crystal with $2$ mW of $405$ nm light emitted from a continuous wave laser diode. Before being coupled into a single-mode fiber, each photon passes through a $10$ nm band-pass filter centered at $810$ nm. The source produces 40,000 pairs per second.

The photons are entangled in polarization because the crystal sits inside a polarizing Sagnac interferometer as shown in Fig. \ref{fig:experiment}. The polarization of the pump is set to $\ket{D}$ so that when pump light hits the two-color ($405$ nm and $810$ nm) polarizing beamsplitter (PBS) that opens the Sagnac, it splits into an equal superposition of illuminating the crystal from the front and back. A two-color half waveplate placed just after the reflected port of the PBS rotates $\ket{V}$ to $\ket{H}$ so that the crystal is illuminated by $\ket{H}$ polarized pump light from both sides. The crystal emits photons in the state $(\ket{HV}\ket{\mathcal{C}} + \ket{HV}\ket{\mathcal{A}})/\sqrt{2}$. The photons in the anti-clockwise path see the two-color half waveplate, rotating their state to $\ket{VH}\ket{\mathcal{A}}$. Then the clockwise and anti-clockwise paths recombine at the two-color PBS and exit the interferometer with the polarization state $(\ket{HV} + e^{i\phi_S}\ket{VH})/\sqrt{2}$, where $\phi_S$ is the relative phase between the two paths in the Sagnac. We apply local polarization rotations on both photons until their state upon exiting the fibers and entering the experiment is as close to $(\ket{HH} + \ket{VV})/\sqrt{2}$ as possible.

\acknow{This work was supported by NSERC and the Fetzer Franklin Fund of the John E. Fetzer Memorial Trust and by grant number FQXi-RFP-1819 from 821 the Foundational Questions Institute and Fetzer Franklin Fund, 822 a donor advised fund of Silicon Valley Community Foundation. 823 A.M.S. is a fellow of CIFAR.}

\showacknow{} 

\bibsplit[13]

\bibliography{main.bib}

\end{document}